\documentclass[lettersize,journal]{IEEEtran}

\usepackage[caption=false,font=normalsize,labelfont=sf,textfont=sf]{subfig}
\usepackage{stfloats}
\usepackage{url}
\usepackage{graphicx}
\usepackage[numbers,sort&compress]{natbib}
\usepackage{tikz}
\usetikzlibrary{trees,decorations}
\usepackage{hyperref}
\usepackage{qcircuit}

\usepackage{amsmath,amsfonts,bm}
\everymath{\displaystyle} 
\usepackage{colortbl}
\usepackage{todonotes} 
\newcommand{\ket}[1]{\ensuremath{\left|#1\right\rangle}} 
\newcommand{\bra}[1]{\ensuremath{\left\langle#1\right|}} 
\newcommand{\braket}[2]{\ensuremath{\left\langle#1|#2\right\rangle}} 

\renewcommand{\bf}[1]{\ensuremath{\mathbf{#1}}}
\newcommand{\norm}[1]
{\ensuremath{\lVert#1\rVert }}
\graphicspath{{figures/}}

\begin{document}

\title{{Dimension reduction and redundancy removal through  successive Schmidt decompositions}

\author{Ammar~Daskin,~\IEEEmembership{}
Rishabh~Gupta,~\IEEEmembership{} 
        Sabre~Kais~\IEEEmembership{}
\thanks{A. Daskin is with the Department
of Computer Engineering, Istanbul Medeniyet University, Istanbul, Turkiye e-mail: adaskin25@gmail.com.}
\thanks{R. Gupta is with the Department of Chemistry,  Purdue  University, West Lafayette, IN, 47907, United States.}
\thanks{S. Kais is with Department of Chemistry, Department of Physics, and Purdue Quantum Science and Engineering Institute, Purdue University, West Lafayette, IN, 47907, United States.}
}

\date{Received: date / Accepted: date}
}


\maketitle

\begin{abstract}
 
Quantum computers are believed to have the ability to process huge data sizes which can be seen in machine learning applications.
In these applications, the data in general is classical. Therefore, to process them on a quantum computer, there is a need for efficient methods which can be used to map classical data on quantum states in a concise manner.
On the other hand, to verify the results of quantum computers and study quantum algorithms, we need to be able to approximate quantum operations into forms that are easier to simulate on classical computers with some errors.

Motivated by these needs, in this paper we study the approximation of matrices and vectors by using their tensor products obtained through successive Schmidt decompositions. 
We show that data with distributions such as uniform, Poisson, exponential, or similar to these distributions can be approximated by using only a few terms which can be easily mapped onto quantum circuits. 
The examples include random data with different distributions, the Gram matrices of iris flower, handwritten digits, 20newsgroup, and labeled faces in the wild. 
And similarly, some quantum operations such as quantum Fourier transform and variational quantum circuits with a small depth also may be approximated with a few terms that are easier to simulate on classical computers. 
Furthermore, we show how the method can be used to simplify quantum Hamiltonians: In particular,  we show the application to randomly generated transverse field Ising model Hamiltonians.
The reduced Hamiltonians can be mapped into quantum circuits easily and therefore can be simulated more efficiently.

\end{abstract}

\section{Introduction}
Quantum computers  are considered to be theoretically more efficient than classical computers: $BPP \subseteq BQP$, which are the complexity classes for the problems that are solvable  with a bounded-error, respectively, on a probabilistic machine and a quantum computer. 
For instance,  in the search for an item among unstructured $N$ items, quantum computers can go beyond the theoretically proven classical lower bound of $O(N)$ and find an item in $O(\sqrt{N})$ steps \cite{grover1997quantum}.  
In addition, although currently there is not a known classical poly-logarithmic algorithm  for integer factorization  or similar problems, these problems can be solved on quantum computers in poly-logarithmic time by using Shor's factoring algorithm\cite{shor1994algorithms} (see Ref.\cite{montanaro2016quantum} for quantum algorithms in general).
However, it is still unknown if $BPP = BQP$ or if the separation is polynomial or exponential (see e.g., \cite{aaronson2022much}). 
Note that although Shor's algorithm indicates an exponential speed-up over the currently known classical algorithms; since the integer factorization is not a proven NP-complete or NP-hard problem,  this does not mean one can solve general NP-complete problems in polynomial time on quantum computers.
However, one can construct some instances similar to (or can be mapped into) the integer factorization and then solve them with Shor's algorithm \cite{pirnay2022super}.
Although we still do not know if those instances naturally exist in any practical problem \cite{szegedy2022quantum}.

Any quantum computation can be represented by an $N$ by $N$ matrix acting on $n = \log N$ number of qubits. 
In general, this matrix requires $O(N^2)$ numbers of single and CNOT gates (or any two-qubit entangling gate) since a matrix may have up to $N^2$ number of independent coefficients \cite{nielsen2010quantum}.
From this, one may argue that simulating a given general matrix computation on quantum computers in $O(poly(n))$ with a certain accuracy may not be possible. 
Therefore, assuming the complexity classes $P$ and  $NP$ are not equal; quantum computers in general cannot solve NP-complete problems with the $n$ number of parameters in $O(poly(n))$ because the solution of these problems can be mapped to a simulating a matrix-vector transform with sizes $O(N^2)$ and $O(N)$ \cite{hillar2013most,pardalos1991quadratic}. 
That means although we know that some problems can be solved more efficiently, it is still not fully known  whether a meaningful quantum computation with $poly(n)$ quantum operations are beyond the capacities of classical computers: i.e., under certain conditions such as a special memory structure; if the result of the computation can be obtained from the given input in $O(poly(n))$ by a classical computer (by using generally randomized algorithms).

The machine learning tasks on quantum computers have started with the various types of proposals for quantum neural networks to either speed-up the learning task or provide a better neural network model: e.g. quantum mechanics modeling human-neuron systems \cite{kak1995quantum},  quantum dots \cite{bonnell1997quantum}, or general quantum computational tools are used to describe classical neural networks \cite{zak1998quantum}.
After Lloyd's seminal paper\cite{lloyd2014quantum} that provided a poly-logarithmic algorithm for principal component analysis of data stored in quantum ram (the data is given as quantum states), the research in quantum machine learning and quantum big data analysis gained a lot of attention and momentum (see  \cite{biamonte2017quantum,schuld2015introduction} for review of the area and \cite{khan2020machine} for an introduction with a comparative analysis to classical machine learning). The quantum principal component analysis leads most researchers to believe that quantum computers may provide exponential speed-up for data analysis tasks. 
However,  it is shown that by using a classical data structure similar to the assumptions made on the quantum algorithm, one can also do the same analysis task on classical computers exponentially faster \cite{tang2019quantum}. 
Although  with the currently known quantum algorithms, exponential speed-up is not known, some quantum versions of the data analysis algorithms are still faster in terms of computation complexity \cite{tang2019quantum}. 
In addition, quantum computers are expected to be able to handle more data, therefore, they would be more powerful in terms of computational space and capacity. 
Some of the quantum versions  of the classical algorithms are also shown to be more accurate for the same machine learning problems \cite{cong2019quantum,chen2022quantum}.  

\subsection{The motivation}
\subsubsection{Mapping data into quantum states to run on quantum computers}
A computation represented as a matrix-vector transformation can be run on quantum computers by first preparing a set of quantum gates that makes the initial state of the qubits equivalent to the vector and secondly by mapping the matrix transformation into known quantum gates. 
The resulting quantum circuit gives the equivalent transformation.
Using parameterized quantum circuits(so-called variational quantum circuits, VQC), one can generate a matrix $U(\bf{\theta})$ whose  transformation depends on the parameters given by the vector $\bf{\theta}$: i.e. the parameters defines the specifics for the quantum gates such as the angle values.
In general,  because of the training and computational difficulties, the design structure $U(\bf{\theta})$ is fixed by using quantum gates which guarantees the entanglement of the qubits and certain computational prowess.

In many recent quantum machine learning models, various types of variational quantum circuits are used as a replacement to a model of neural network. Given different inputs as the initial state of the quantum circuit, training in these models is generally done by classically optimizing the parameters of the circuits.
In other words, the learning task aims to find a circuit $U(\bf{\theta})$ so that for an input $\ket{\psi}$ representing a data vector, $U(\bf{\theta})\ket{\psi}$ outputs a state from which the expected result can be obtained through quantum measurements with high probability.

In this learning model, one of the fundamental tasks is to find a way to map a data vector $\bf{x}$ into a quantum state \ket{\psi} without losing any information. 
This can be done in two different ways \cite{schuld2015introduction,biamonte2017quantum, huang2021power,daskin2022walk}: i) as done in neural networks, we use one qubit for each feature $x_i$.
Because of the similarities to classical neural networks, this mapping provides a natural way to do quantum machine learning with variational quantum circuits. However, the difficulty is that the required number of qubits may be very high and is very likely beyond the capabilities of current (may be also near future) quantum computers.
ii) We can consider $\ket{\psi} =  \bf{x}/\norm{\bf{x}}$. 
Although the learning task may become more difficult, the number of qubits becomes exponentially small in comparison to the dimension of the data vector, we will use $N=2^n$.
However, generating a general $\bf{x}$ requires $O(N)$ quantum gates. 
This complicates the simplicity of the variational quantum circuits and impedes the applications to the problems with high dimensional data vectors.
As a result, as in the classical data analysis, for quantum computers we need to find methods similar to the singular value decomposition \cite{golub2013matrix}  to reduce the dimensions of data sets in a way so that we can reduce the required number of qubits and the number of quantum gates.

\subsubsection{Classical simulation of quantum circuits and designing more efficient quantum circuits}
Simulating quantum operations is difficult when the qubits are entangled.
It is known that quantum operations can be efficiently simulated by taking advantage of their mathematical structures which are generally studied by using tensor networks \cite{biamonte2017tensor,biamonte2019lectures}. 
It is also well known that the bipartite entanglement can be understood by using the Schmidt decomposition of the bipartite system \cite{terhal2000schmidt,kais2007entanglement}.

 Recently, it is shown that the coefficients in the Schmidt decomposition of quantum Fourier transform exponentially decays; thus, the quantum Fourier transform may be efficiently simulated on classical computers and also may be more efficient than the classical algorithms for the discrete Fourier transform \cite{chen2022quantum}. 
It is also shown that the entanglement can be classically  forged through the Schmidt decomposition  to combine the results of two separate quantum operations  on classical computers \cite{eddins2022doubling}.

Motivated by the above two points, in this paper we study the approximation of matrices and vectors by using their tensor products obtained through successive Schmidt decompositions. 
We show that some data distributions can be approximated by using a few terms which can be easily mapped into quantum circuits. 
Similarly, some quantum operations also may be approximated with high accuracy that are easier to simulate on classical computers.

In the following, we first explain Schmidt decomposition and draw an equivalent recursion tree that can be used to understand the decomposition. 
Then, we describe how this tree can be used to eliminate the terms with lower coefficients and generate approximated quantum operators.
In Sec.\ref{Sec:approximations}, we show the applications of the method to random data vectors, Gram matrices, and general symmetric matrices, example machine learning datasets: Iris flower, handwritten digits, 20newsgroup and labeled faces in the wild. Then, after discussing how it can be used in solving systems of linear equations, we show the approximation of quantum Fourier transform in Sec.\ref{Sec:qft}.
Then, we study the dataset with a circular type distribution in Sec.\ref{Sec:circular} and variational quantum circuits with different depths in Sec.\ref{Sec:vqc}.
In our final example, in Sec.\ref{Sec:hamiltonian} we show how the method can be applied as an approximation tool to reduce the terms in transverse field Ising model quantum Hamiltonians.   
After showing how classical and quantum computation can be done with the approximated forms, we conclude the paper.
\section{The Schmidt Decomposition and its recursion tree structure}
In the following notations, we use bold face or the Dirac bra-ket notations for the vectors: \ket{x} and \bf{x} are equivalent, but the former represents a quantum state which is in general normalized. The matrices are written in capital letters.

\subsection{Singular value decomposition (SVD)\cite{golub2013matrix}}

    For any $M\times N$ matrix $A$, there exists a factorization of the form:
    \begin{equation}
     \label{Eq:svd}
        A=U\Sigma V^* = \sum_i^r \sigma_i \bf{u_i} \bf{v_i^*} 
    \end{equation}where $U$ and $V$ are $M\times M$ and $N\times N$ unitary matrices and $\Sigma$ is an $M\times N$ rectangular diagonal matrix with positive diagonal entries $\sigma_1 \geq \sigma_2 \dots \geq 0$, known as singular values. $r$, the number of non-zero singular values, defines the rank of the matrix and is min of $M$ and $N$.

    Let $vec(A)$ be the vectorized form of the matrix $A$. 
    The SVD can be used to write this vector into a sum of weight-ordered separable components: 
    \begin{equation}
    \label{Eq:svd_vector}
    vec(A)=\sum_i\sigma_i \bf{u_i} \otimes \bf{v_i}    
    \end{equation}

 Note that SVD provides the closest low-rank approximations in $l^2$-norm to the matrix: For instance a rank $r_1\leq r$ approximation can be obtained by:
 \begin{equation}
 A = \sum_i^{r_1} \sigma_i \bf{u_i} \bf{v_i^*} 
 \end{equation}
 
\subsection{Schmidt decomposition\cite{terhal2000schmidt,kais2007entanglement}}
Let \ket{\psi} represent an $n$-qubit system ($N=2^n$), where qubits are ordered as $q_0q_1\dots q_{n-1}$. 
Using the Schmidt coefficients between the first qubit $q_0$ and the rest of the system: i.e., $q_0\rule[1pt]{5mm}{.4pt}(q_1\dots q_{n-1})$, we can represent the state in the following form:
\begin{equation}
    \ket{\psi} = \sigma_1 \ket{u_1}\otimes \ket{v_1}+ \sigma_2 \ket{u_2}\otimes \ket{v_2}.
\end{equation}
In the above, while $\sigma_1^2 + \sigma_2^2= 1$, $\braket{u_1}{u_2}=1$  and $\braket{v_1}{v_2}=1$. 
The Schmidt decomposition of a vector can be found by converting the vector into a $2 \times 2^{n-1}$ matrix and using the SVD described above. In that case, $\sigma_1$ and $\sigma_2$ are the singular values and $\ket{u_i}$ and $\ket{v_i}$s are the singular vectors.

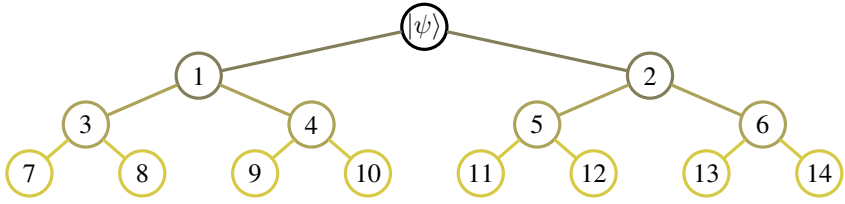
\begin{figure*}[ht]
    \centering
    \tikzstyle{every node}=[fill]
\tikzstyle{edge from parent}=[decoration=lineto,segment length=5mm,
                              segment angle=160,draw]
\begin{tikzpicture}[
scale=0.5, very thick, every node/.style = {draw, circle, 
minimum size = 6mm, 
inner sep=0pt, 
text centered,
  text=black},
    grow = down,  
    level 1/.style = {sibling distance=12cm},
    level 2/.style = {sibling distance=6cm}, 
    level 3/.style = {sibling distance=3cm}, 
    level distance = 1.25cm,
    shape=circle,very thick,level distance=13mm,
                    cap=round]
\node[text=black] {$\ket{\psi}$} 
child [color=black!60!yellow]{
    node{1} child [color=black!40!yellow] 
    {
        node{3}child [color=black!20!yellow] {node{7}} 
        child [color=black!20!yellow] {node{8}}
    }
    child [color=black!40!yellow] 
    {
        node{4}child [color=black!20!yellow] {node{9}} 
        child [color=black!20!yellow] {node{10}}}
   }
child [color=black!60!yellow]{
    node{2} child [color=black!40!yellow] 
    {
        node{5}child [color=black!20!yellow] {node{11}} 
        child [color=black!20!yellow] {node{12}}
    }
    child [color=black!40!yellow] 
    {
        node{6}child [color=black!20!yellow] {node{13}} 
        child [color=black!20!yellow] {node{14}}}
    };
\end{tikzpicture}
    \caption{The recursion tree for the Schmidt decomposition of \ket{\psi}: A node with label $i$ represents the term $\sigma_i\ket{u_i}\ket{v_i}$. 
    The sum of the child-terms gives the Schmidt decomposition for the vector $v_{i}$ in the parent node $i$.}
    \label{fig:rtree}
\end{figure*}

Here,  we can keep recursively taking the Schmidt decomposition of the larger vectors, i.e. \ket{v_i}s, and obtain the following: 
\begin{equation}
\begin{split}
    &\ket{v_1} = \sigma_3 \ket{u_3}\otimes \ket{v_3}+ \sigma_4 \ket{u_4}\otimes \ket{v_4},
    \text{ and }\\
 &\ket{v_2} = \sigma_5 \ket{u_5}\otimes \ket{v_5}+ \sigma_6 \ket{u_6}\otimes \ket{v_6}.
\end{split}
\end{equation}
Note that the size of any $\ket{u_i}$ is still 2 by 1. Once the size of $\ket{v_i}$ also goes down to 2, we stop the recursion.
In the final analysis, we can write \ket{\psi} in the following decomposed form:
\begin{equation}
\label{eq:terms}
\begin{split}
\ket{\psi} = &
    \sigma_1 \ket{u_1}\otimes \sigma_3 \ket{u_3} \otimes \sigma_7 \ket{u_7}\otimes\dots \\
    &+
    \sigma_1 \ket{u_1}\otimes \sigma_3 \ket{u_3} \otimes \sigma_8 \ket{u_8}\otimes\dots \\
    &+\\
    & \vdots \\
    &+
    \sigma_2 \ket{u_2}\otimes \sigma_6 \ket{u_6} \otimes \sigma_{14} \ket{u_{14}}\dots\\
    &+\dots    =  \sum_i \ket{\psi_i}  
\end{split}
\end{equation}
where $\ket{\psi_i}$ represents a term in the summation and the number of terms is exponential in the number of qubits.
Fig.\ref{fig:rtree} shows the recursion tree for the above equation. In the figure, each path from the root node to a leaf node represents a summation term in the  equation. If any of the coefficients $\sigma_i$ is zero in this path, then the equivalent term in the summation is  zero as well.
Therefore, the number of non-zero terms is equal to the number of paths with non-zero coefficients.

\subsection{Quantum Operations}
The same recursion tree can be generated also for the operators. 
In that case, we use the following steps:
\begin{itemize}
    \item Let $A$ be an N by N matrix. First, we vectorize the matrix A (Column- or row-based vectorization can be used. In this paper we will assume row-based vectorization.):  
    \begin{equation}
        \ket{\psi} = vec(A).
    \end{equation} 
    \item Then we draw the recursion tree as in Fig.\ref{fig:rtree}.
    \item Since each path from the root to a leaf node is an $N^2$ by 1 vector, we can convert these terms back to $N$ by $N$ matrices.
    In that case, we can write $A$ as in the form:
    \begin{equation}
    \label{eq:SchmidtFormA}
        A = \sum_{i\in \{0,\dots, N^2/2\} | A_i \neq 0} A_i
    \end{equation}
    \item Note that the number of paths is equal to the number of leaf nodes which is $2^{n}/2$ for a vector of dimension $2^n$.
    \item To convert a tensor decomposition of a vector into an operator in tensor form, let us consider the following example term: 
    \begin{equation}
        \ket{\psi_j} =  \ket{p_1} \otimes \ket{p_k}\otimes \ket{q_1}\otimes  \dots \otimes\ket{q_k}
         \label{Eq:psij}
    \end{equation}
    Assuming \ket{p_i} and \ket{q_i}s are column vectors of dimension 2; from the definitions in Eq.\eqref{Eq:svd} and \eqref{Eq:svd_vector}, we can convert this term into an operator $A_j$ as follows: 
    \begin{equation}
        A_j =  \ket{p_1}\bra{q_1} \otimes \ket{p_2}\bra{q_2}\otimes \dots \otimes \ket{p_k}\bra{q_k} = \bigotimes_{i=1}^k \ket{p_i}\bra{q_i}. 
        \label{Eq:Aj}
    \end{equation}
    Here, each \ket{p_i}\bra{q_i} is a 2 by 2 matrix and is not unitary. However, they can be written as a sum of two unitary matrices.
\end{itemize}

\section{Approximation by removing the number of paths with smaller coefficients}
\label{Sec:approximations}
Choosing a threshold probability in the tree, we can approximate any operator or vector by eliminating the paths with coefficients lower than the threshold probability.
To determine the approximation error between the normalized quantum state \ket{\psi} and the approximation \ket{{\phi}} (not-normalized), we shall use the norm of the difference: 
\begin{equation}
   \epsilon = \left\lVert\ket{\psi} - \ket{\phi} \right\rVert_2.
\end{equation}
Note that this is equivalent to the sum of squared errors. One can also consider the mean squared error: $\epsilon^2/2^n$, which is much smaller.
In the following subsections, we show approximations of different cases and their approximation errors.

\subsection{Approximation of Gram Matrix}
\begin{figure*}[ht]
    \centering
    \subfloat{\includegraphics[width=3.5in]{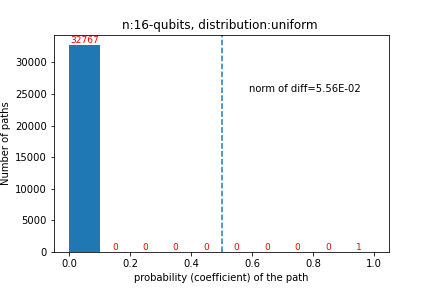}}\hfill
    \subfloat{\includegraphics[width=3.5in]{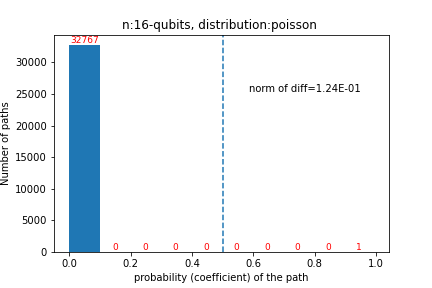}}\\
    \subfloat{\includegraphics[width=3.5in]{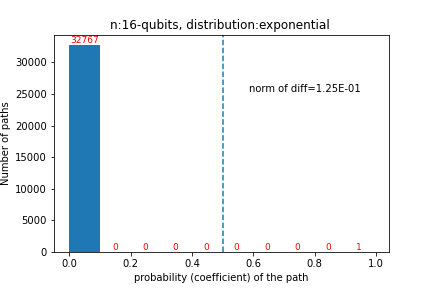}}\hfill
    \subfloat{\includegraphics[width=3.5in]{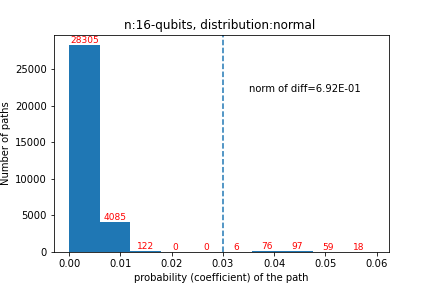}}
    \caption{Histogram of the coefficients for  $G$ produced by randomly generated $X$  with different distributions. $n$ represents the number of qubits required for $vec(G)$: i.e. the dimension of $vec(G)$ is $2^n$. }
    \label{fig:histofgram}
\end{figure*}

Gram matrix\cite{golub2013matrix} is frequently  used in machine learning and other areas \cite{shawe2005eigenspectrum,ramona2012multiclass,sastry2020detecting}. For a given data matrix $X$ (the column vectors represent the data),  it is defined by the following product:
\begin{equation}
    G = X^TX.
\end{equation}
The matrix $G$ is  positive semi-definite and its rank depends on the number of independent data vectors in $X$. 
In the simulations, we have used the normalized $vec(G)$ produced from the default random number generator used in the Python-numpy package. 
Using different distributions with default parameters, an example of the output which shows the histogram of the probabilities of the paths is presented in Fig.\ref{fig:histofgram} (Most output generates a similar result, therefore we only show one instance for each to explain the behavior in each distribution.). 
As seen in the figure,  in the cases of the exponential, Poisson, and uniform distributions, the matrix $G$ can be approximated with high accuracy by using only one term. 
This is because the Schmidt decomposition is related to the eigenvalues of the data matrix and the variance of the eigenvalues in these cases is large: i.e. there is one dominant eigenvalue.
When the distribution is normal although there is a clear cut-off point, the approximation accuracy is less than the others. 

\begin{figure*}[ht]
    \centering
    \subfloat{\includegraphics[width=3.5in]{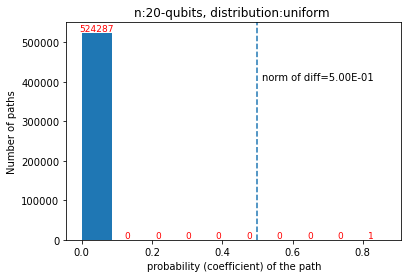}}\hfill
    \subfloat{\includegraphics[width=3.5in]{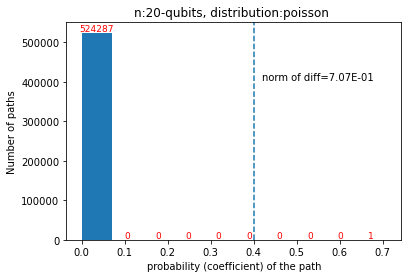}}\\
    \subfloat{\includegraphics[width=3.5in]{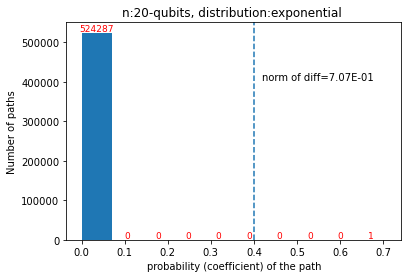}}\hfill
   \subfloat{ \includegraphics[width=3.5in]{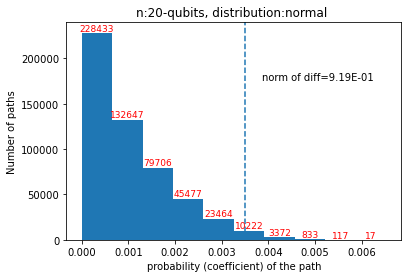}}
    \caption{Histogram of the coefficients for  $X$ produced randomly  with different distributions. $n$ represents the number of qubits required to represent $vec(X)$: i.e. the dimension of $vec(X)$ is $2^n$. }
    \label{fig:histofrandX}
\end{figure*}
Here note that one can also use the same technique to directly approximate $X$. 
Fig.\ref{fig:histofrandX} shows the direct approximations for instances of $\ket{\psi}=vec(X)$ generated randomly from different distributions. As it can be seen from the figure, although a cut-off point  still exists in all except the normal distribution, the accuracy is much lower.

\subsubsection{Applications in data science}
\begin{figure*}[ht]
    \centering
    \subfloat{\includegraphics[width=3.5in]{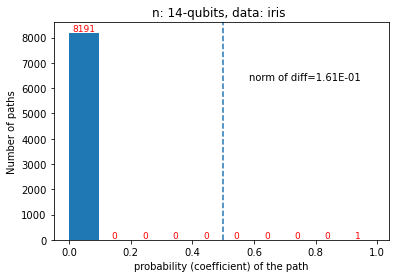}}\hfill
    \subfloat{\includegraphics[width=3.5in]{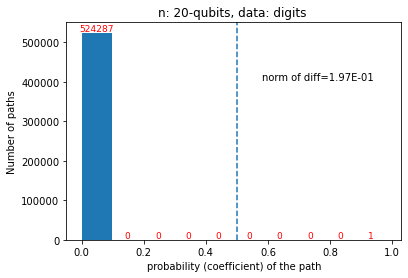}}\\
    \subfloat{\includegraphics[width=3.5in]{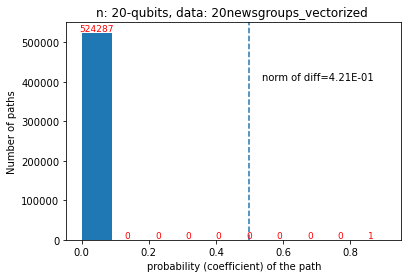}}\hfill
     \subfloat{\includegraphics[width=3.5in]{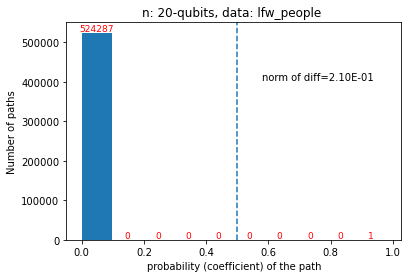}}
    \caption{Histogram of the coefficients for  $G$ produced by an example dataset $X$. $n$ represents the number of qubits required for $vec(G)$: i.e. the dimension of $vec(G)$ is $2^n$, which is related to the number of samples (chosen to be the power of 2) and the number of features. }
    \label{fig:histofmldata}
\end{figure*}

We also find the kernel matrix $G$ for example real-world datasets \cite{Dua:2019} used in machine learning algorithms: Iris flower data, images of handwritten digits data, vectorized form of 20 newsgroups data, and labeled faces in the wild\cite{LFWTechUpdate} data which are loaded  through Python scikit-learn package \cite{sklearn_api}. 
To make the size of $G$ a power of 2, we choose the first 128 samples for the iris dataset, and 1024 samples for the other datasets. As seen in Fig.\ref{fig:histofmldata}, the norm of the approximation error is 0.161 for the iris, 0.421 for the newsgroup, 0.197 for the digits, and 0.21 for the faces. 
Here, note that the norm gives us the accumulated error, the mean squared error is much smaller: e.g. for the faces, it is around $0.42x10^{-7}$. 

These results indicate that in many cases we can approximate $G$ as a single term in the following form: 
$\alpha Q_{1} \otimes \dots \otimes Q_{n}$ or as a sum of a few  of these terms. 
As shown in Sec.\ref{Sec:classicalcomputation} and Sec.\ref{Sec:quantumcomp}, these reduced forms can be used to do classical and quantum computations more efficiently.

\subsubsection{Applications to solve systems of linear equations}
A system of linear equations \cite{saad2003iterative} is in general defined by $\bf{y}:= A\bf{x}$.
In the theoretical sense, the solution of this equation can be found by computing the inverse  $A^{-1}$: i.e. $\bf{x} = A^{-1}\bf{y}$ (In practice, this would not be the most efficient way.). 
In some cases, since $A$ may be a singular matrix, instead of solving the original equation, the equation is converted into the following normal equation: $A^T\bf{y} := A^TA\bf{x}$. Then, the solution is found by the inverse $(A^TA)^{-1}$.  Here, $G = A^TA$ is the same as the Gram matrix generated by the column vectors of $A$. Therefore, the analysis in Fig.\ref{fig:histofgram} is also the same for this matrix. 
If this matrix can be approximated with one term: $G\approx \alpha Q_{1} \otimes \dots \otimes Q_{n}$, then the inverse  can be found by the following: 
\begin{equation}
    G^{-1}\approx \alpha^{-1} Q_{1}^{-1} \otimes \dots \otimes Q_{n}^{-1}.
\end{equation}
The inverse in this form would be directly represented since each $Q_i$ is of dimension 2.

\subsection{Approximation of the symmetric matrix $(X + X^T)$}
Note that the arguments for the Gram matrix are also applicable to the symmetric matrix $(X + X^T)$, assuming $X$ is a square real matrix. That means if the distribution of the data is uniform or similar, the tensor approximation can be used with high accuracy for these cases either.

\subsection{Approximation of the quantum (or discrete) Fourier transform}
\label{Sec:qft}
The quantum Fourier Transform (QFT) is the quantum version of the discrete Fourier transform. For $w=e^{2\pi i/N}$ and $M=(N-1)$, it can be written in the following unitary matrix form \cite{nielsen2010quantum}:
\begin{equation}
    QFT = \frac{1}{\sqrt{N}} \left(\begin{matrix}
1&1&1&\cdots &1 \\
1&\omega&\omega^2&\cdots&\omega^{M} \\
1&\omega^2&\omega^4&\cdots&\omega^{2M}\\ 
1&\omega^3&\omega^6&\cdots&\omega^{3M}\\
\vdots&\vdots&\vdots&&\vdots\\
1&\omega^{M}&\omega^{2M}&\cdots&\omega^{M^2}
\end{matrix}\right)
\end{equation}
QFT can be implemented in logarithm time on quantum computers: i.e. $O(log(N)^2)$. 
Therefore, it can be employed to provide exponential efficiency over the classical algorithms as done in Shor's integer factorization\cite{shor1994algorithms}. 
As mentioned in the introduction, it is shown that quantum Fourier transform can be approximately decomposed into two equal dimension local parts( as in $QFT = A\otimes B$) because the Schmidt coefficients decay exponentially with the size \cite{chen2022quantum}.

\begin{figure*}[ht]
    \centering
     \subfloat{\includegraphics[width=3.5in]{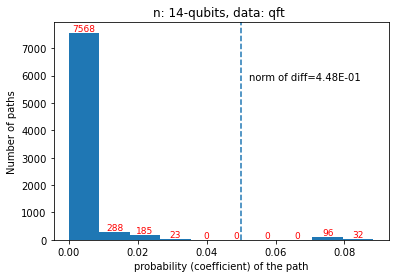}}\hfill
     \subfloat{\includegraphics[width=3.5in]{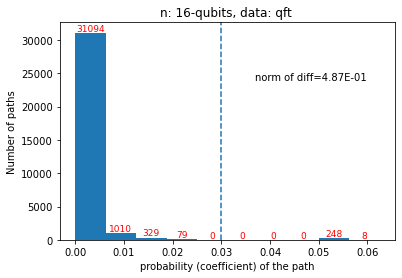}}\\
    \subfloat{\includegraphics[width=3.5in]{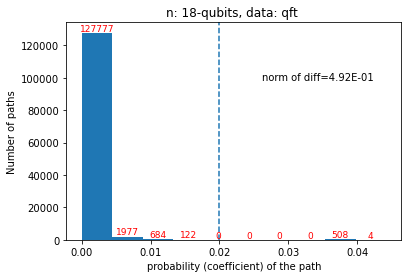}}\hfill
    \subfloat{\includegraphics[width=3.5in]{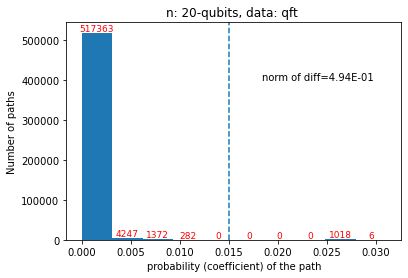}}
    \caption{Histogram of the coefficients for QFT: Here, $n$ represents the number of qubits required for $vec(QFT)$: i.e. the dimension of $vec(QFT)$ is $2^n$. Therefore, the sizes of the QFT matrices are of dimensions $2^7, 2^8, 2^9, 2^{10}$.  }
    \label{fig:qft}
\end{figure*}

We also use our method to decompose $vec(QFT)$ as shown in Fig.\ref{fig:qft}. 
From the figure, we can see that there is a clear cut-off point to make an approximation. And a careful observation reveals that the number of terms on the right-hand side of the figure (the terms with larger coefficients) grows linearly with the matrix dimension. 
That means the vector $vec(QFT)$ can be approximated by using $N$ terms. 
Then by using Eq.\eqref{Eq:psij} and \eqref{Eq:Aj}, the matrix $QFT$ requires $N/2$ terms decomposed into a tensor of 2 by 2 matrices. 

\subsection{Data having a type of circular distributions}
\label{Sec:circular}
\begin{figure*}[ht]
    \centering
     \subfloat{\includegraphics[width=3.5in]{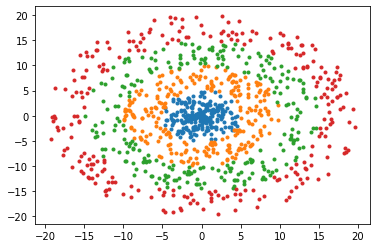}}\hfill
      \subfloat{\includegraphics[width=3.5in]{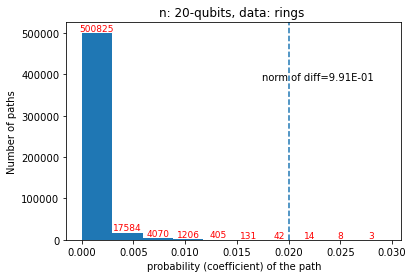}}
    \caption{Rings and its histogram of the coefficients. There is no clear cut-off point for the approximation and the approximation error is higher. }
    \label{fig:rings}
\end{figure*}

In Fig.\ref{fig:histofgram}, we have seen that in the normal distribution, although there is a clear partition point, the approximation accuracy is less than the others. In circular distributions, in general, we did not observe a clear cut-off point. 
For instance, consider the rings in Fig.\ref{fig:rings}, although the number of paths with the larger coefficients seems to be much less than others, there is no clear cut-off point for the approximation and the coefficients are in general very close to each other.

A similar pattern can be also observed in random unitary matrices which can be seen in the following section.

\subsection{Approximation of the variational quantum circuits}
\label{Sec:vqc}
\begin{figure}[ht]
    \centering
    \input{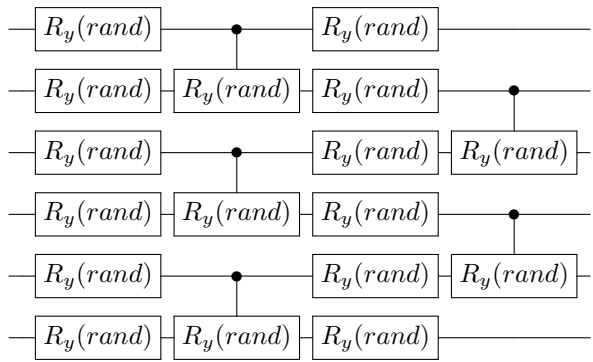}
    \caption{A variational quantum circuit (VQC) with depth 4. $rand$ indicates that each gate is generated with a random angle. }
    \label{fig:vqc}
\end{figure}

Variational quantum circuits are parameterized circuits that include some single and control (entangling) gates.
In our example circuit, to work with the real entries, we only use the following rotation about the-y axis:
\begin{equation}
R_y(\theta) = \left(\begin{matrix}
    cos(\theta/2) & sin(\theta/2)\\
    -sin(\theta/2) & cos(\theta/2 )
\end{matrix}\right),
\end{equation}
where $theta$ is an angle value. We generate this angle randomly (by using normal distribution) for each quantum gate in our circuit which is shown  in Fig.\ref{fig:vqc}.
The depth of a quantum circuit is defined as the largest number of quantum gates along any line (each line represents a qubit).
Thus, the circuit in Fig.\ref{fig:vqc} is of depth 4.
In the numerical simulations, we use the depth values 4, 8,12, and 16 which basically is the same as the drawing in Fig.\ref{fig:vqc} repeatedly required 1, 2, 3, and 4 times, respectively.

\begin{figure*}[ht]
    \centering
     \subfloat{\includegraphics[width=3.5in]{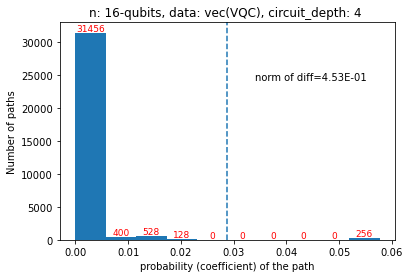}}\hfill
     \subfloat{\includegraphics[width=3.5in]{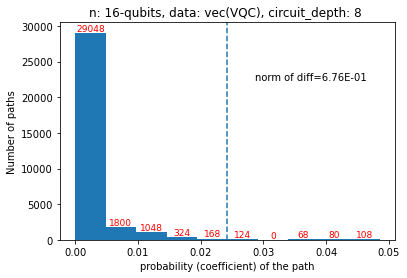}}\\
    \subfloat{\includegraphics[width=3.5in]{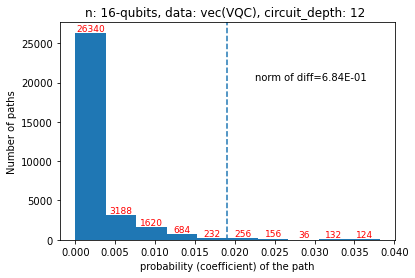}}\hfill
    \subfloat{\includegraphics[width=3.5in]{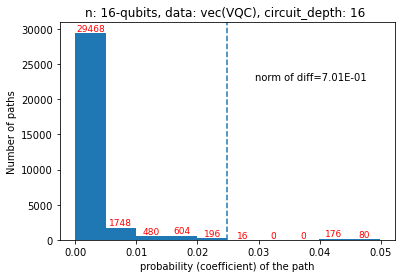}}
    \caption{Histogram of the coefficients for VQC with different depth values: Here, $n$ represents the number of qubits required for $vec(VQC)$. The error grows with depth of the circuit. }
    \label{fig:histofvqc}
\end{figure*}

The histograms of the coefficients are shown in Fig.\ref{fig:histofvqc} where the cut-off point is chosen as the middle point. As seen in the figure, the approximation errors grow with the depth of the circuit.
This is because the matrix starts to depend on more parameters and  to look like more of a random unitary matrix with random matrix elements. This becomes similar to a circular type distribution and as seen in the previous chapter, the method would not produce a good result for these types of data matrices.

\subsection{Approximation of Hamiltonians} \label{Sec:hamiltonian}
\begin{figure*}[ht]
    \centering
     \subfloat{\includegraphics[width=3.5in]{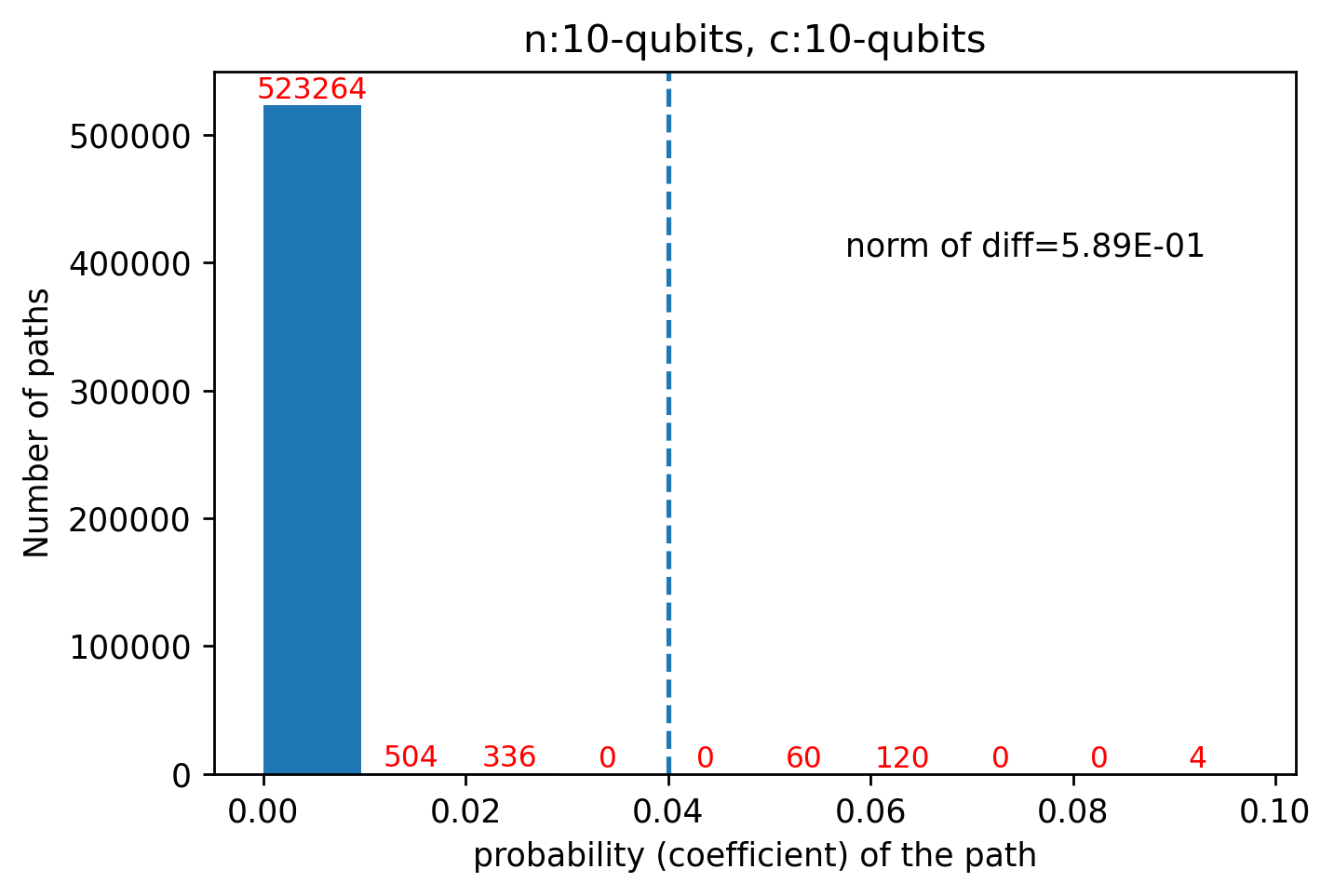}}\hfill
     \subfloat{\includegraphics[width=3.5in]{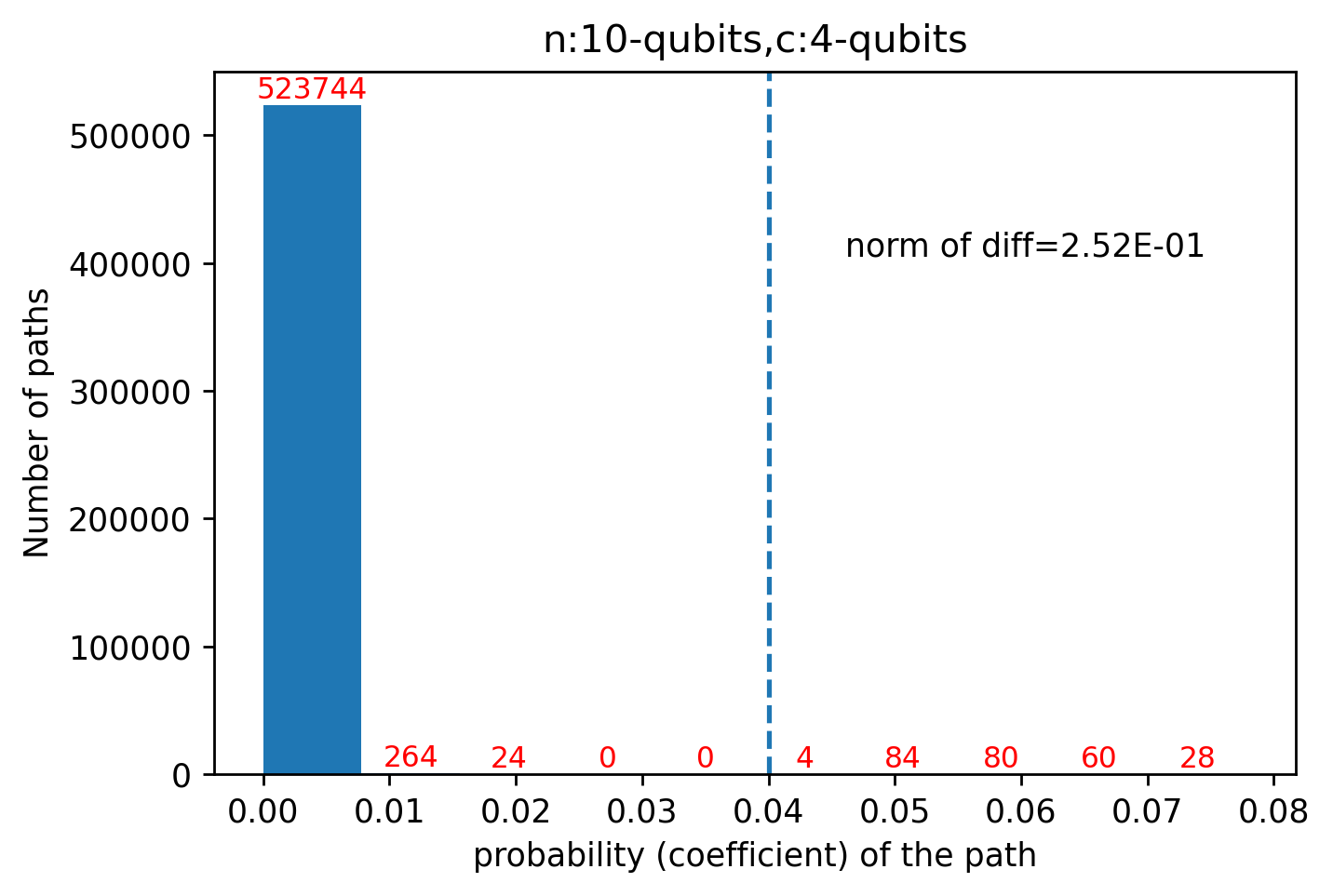}}
    \caption{Histogram of the Schmidt coefficients corresponding to TFIM Hamiltonian of a 10-qubit quantum system with \textit{c}-qubits that are coupled. With fewer coupling terms we see that the norm of difference reduces for the same cut-off probability.}
    \label{fig:histofhamil}
\end{figure*}

\begin{figure*}[ht]
    \centering
     \subfloat{\includegraphics[width=3.5in]{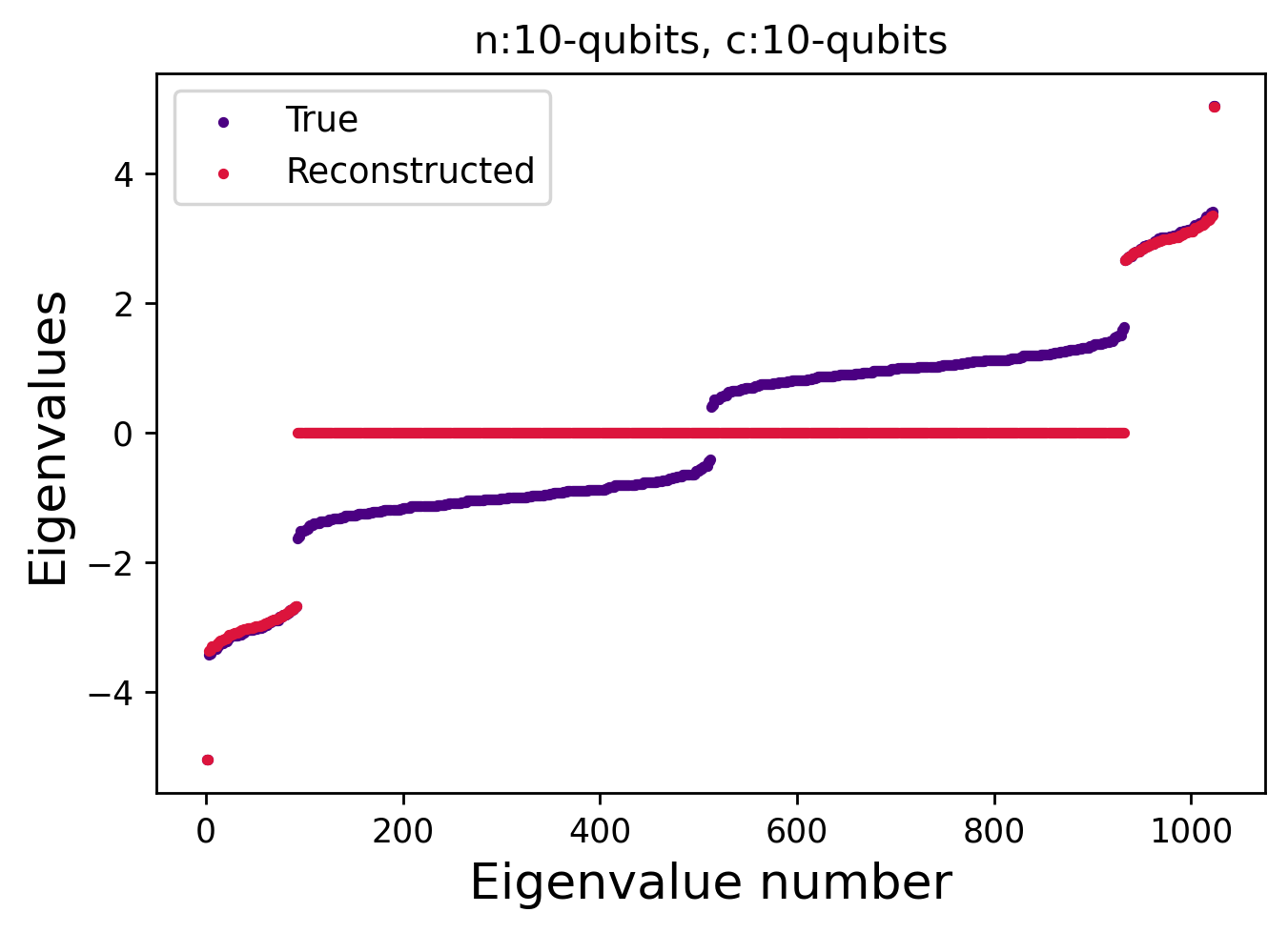}}\hfill
     \subfloat{\includegraphics[width=3.5in]{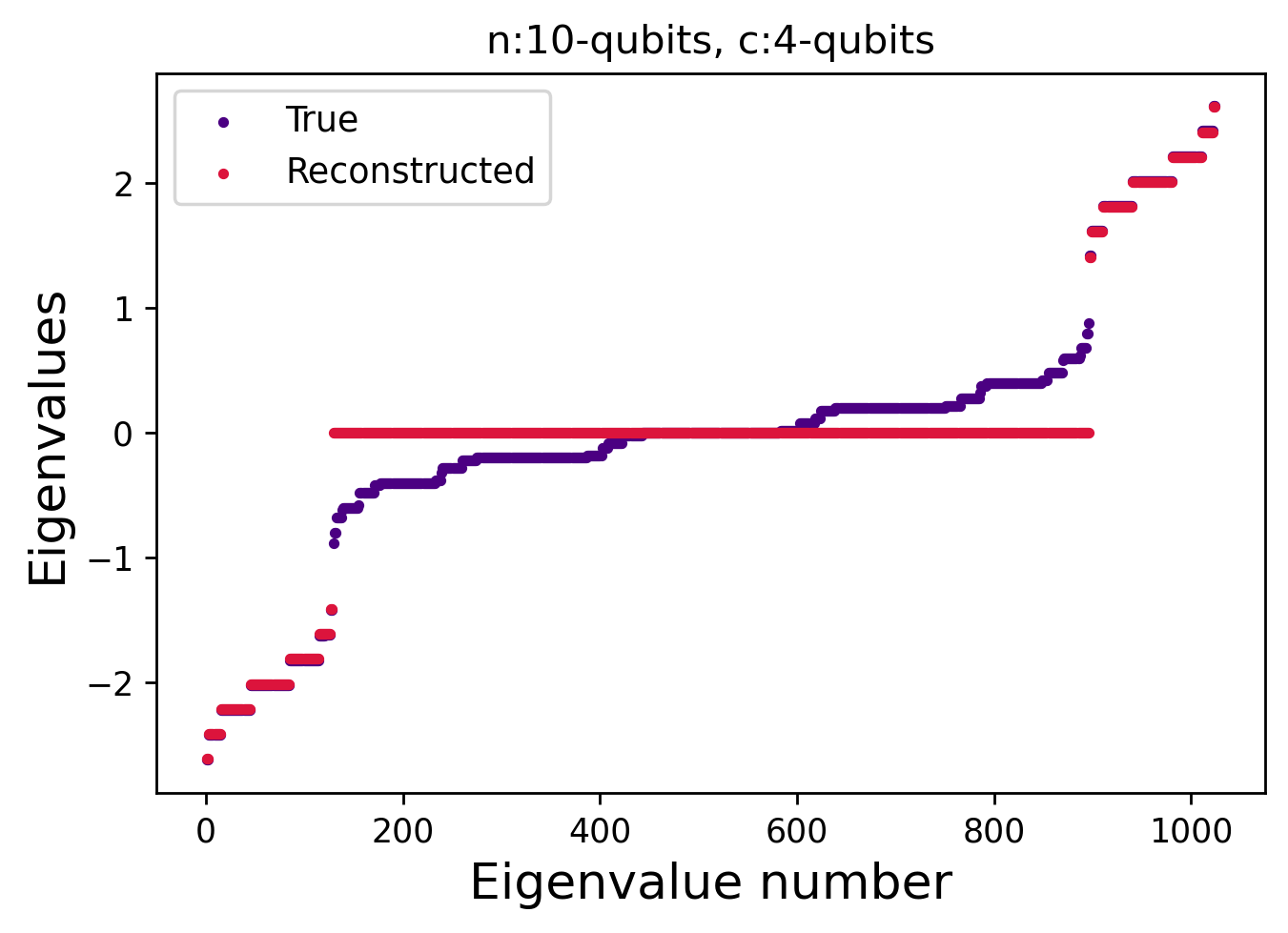}}
    \caption{Eigenvalue spectrum of the true and the reconstructed TFIM Hamiltonian for a 10-qubit quantum system. As can be seen that the Hamiltonian approximation using our proposed Schmidt decomposition protocol is able to reproduce ground state and excited state eigenvalues with very good accuracy.  }
    \label{fig:histofhamil_eval}
\end{figure*}

In quantum mechanics, the Hamiltonian of a system is a key quantity that determines the time evolution of a quantum state. The eigenvalue spectrum of the Hamiltonian provides important information about the system's behavior such as its stability, spectral gaps, symmetries of system, and the transitions between energy levels that are possible. There are various techniques that try to characterize quantum systems by understanding their full Hamiltonian through various tomographic techniques \cite{yu2022practical, haah2021optimal, krastanov2019stochastic, evans2019scalable, bairey2019learning, qi2019determining, gupta2022hamiltonian, gupta2021maximal, gupta2021convergence, gupta2022variation}. However, as the size of the quantum systems increases calculating the full eigenvalue spectrum can be computationally expensive, and a limited number of eigenvalues may be sufficient. For example, in quantum algorithms such as quantum Monte Carlo, it may only be necessary to know the lowest eigenvalue of the Hamiltonian and not the full spectrum \cite{huggins2022unbiasing}. In some optimization problems, such as the calculation of the ground state energy of a system, determining the lowest eigenvalue may be only sufficient and not the entire spectrum. And therefore, approximating the Hamiltonian through successive Schmidt decomposition can prove to be a useful tool for addressing such problems and also for tackling large quantum systems. \\
To test our proposed approach for Hamiltonian approximation, we considered the Transverse Field Ising Model (TFIM) Hamiltonian, defined by $H= h\sum_{i}^{n}\sigma_x^i + J\sum_{i,j=i+1}^{c}\sigma_z^i\sigma_z^j$ where \textit{h} and \textit{J} are the transverse field and coupling parameters, respectively, \textit{n} is the total number of qubits, and \textit{c} is the number of qubits that are coupled. For computing the results we choose \textit{h} = 0.1 and \textit{J} = 0.5 for a 10-qubit quantum system. Figure \ref{fig:histofhamil} shows the histogram of the coefficients with a cutoff probability of 0.04 for which the norm of difference is 0.589 for the system where all spins are coupled and 0.252 for the system where only 4 spins are coupled. The interesting thing about such an approximation can be seen in the eigenvalue distribution of the true and the reconstructed Hamiltonian in Figure \ref{fig:histofhamil_eval}. Approximating the Hamiltonian using only a few Schmidt coefficients that are large we are able to generate the eigenvalue spectrum that accurately reproduces the eigenvalues that are not close to 0. This is because if we do the SVD of a Hamiltonian matrix, the singular values correspond to the magnitude of the eigenvalues, and since we ignore the Schmidt coefficients that are small, the reconstructed Hamiltonian has information of the eigenvalues that are away from 0 as is seen in Figure \ref{fig:histofhamil_eval}. 

\section{Computations with the Schmidt forms }
\subsection{Classical computation}
\label{Sec:classicalcomputation}
After writing a matrix $A$ in the form of Eq.\eqref{eq:SchmidtFormA} and a vector \ket{\psi} in the form of \eqref{eq:terms}, one can do the matrix-vector transforms in this forms. 
For any $i, j$, assume that the decomposition of $A_i$ and \ket{\psi} is given as $A_i =\alpha Q_{1} \otimes \dots \otimes Q_{n}$ and $\ket{\psi_1} = \beta \ket{p_1} \otimes \dots \otimes \ket{p_n}$, where $\alpha, \beta$ are real coefficients and $Q_k$s and \ket{p_k}s are matrices and vectors of dimension 2.
The product $A_i \ket{\psi_j}$  can be found as:
\begin{equation}
    A_i \ket{\psi_j} = \alpha\beta Q_{1}\ket{p_1} \otimes \dots \otimes Q_{n}\ket{p_n}
\end{equation}
Here, any $Q_k\ket{p_k}$ requires only 4 addition and multiplication, therefore one can generate the above equation in tensor form in $O(n)$ time and generating the whole vector requires $O(2^n)$ operations and memory. However, a single entry can be computed without generating the whole vector. Therefore, a single entry can be obtained in $O(n)$ time.

If $A$ is approximated by using $r$ number of $A_i$s, then $A\ket{\psi_j}$ can be found in $O(r2^n)$ times. However, if one only interested in a single entry of the output vector $A\ket{\psi_j}$, it can be found in polynomial time, i.e. $O(rn)$ time, without computing the whole vector. 

As seen in the case of QFT, there were $N/2$ terms. That means by this approximation, an entry of $QFT\ket{\psi_j}$ can be approximated in $O(\frac{N}{2}n)$ time.
\subsection{Quantum computation}
\label{Sec:quantumcomp}
Assuming $A = \sum_i A_i$ and each $A_i$ is a unitary matrix. A matrix given as a sum of unitary matrices can be implemented as quantum circuits by using different control bits to control each term \cite{childs2012hamiltonian,daskin2012universal,berry2015simulating,daskin2019context}. 
If the individual circuit for each term includes the same quantum gates, they can be used to reduce the number of quantum gates in the overall circuit \cite{daskin2012universal,daskin2014quantum}.

Since each $A_i$ is a tensor decomposition of 2 by 2 matrices in the form: $A_i =\alpha Q_{1} \otimes \dots \otimes Q_{n}$. 
Each term requires only $O(n)$  quantum gates. Note that if $Q_j$s  are non-unitary matrices, then $A_i$ can be considered as a sum of two unitary matrices each of which in tensor form. That means we write each $Q_j$ as a sum of unitary matrices (see Ref.\cite{daskin2019context} for a similar concept). 
As an example the following can be considered as the circuit equivalent of the summation of two terms, $\alpha Q_{1} \otimes Q_2 + \beta Q_{3}\otimes Q_4$:
\begin{center}
\centering
\mbox{
     \Qcircuit @C=0.5em @R=0.5em {
&	\gate{C_1}	&	\ctrlo{1}	&	\ctrlo{2}	&	\ctrl{1}	&	\ctrl{2}	&	\gate{C_2}	\qw	\\
&	\qw	&	\gate{Q_1}	&	\qw	&	\gate{Q_3}	&	\qw	&	\qw	\qw	\\
&	\qw	&	\qw	&	\gate{Q_2}	&	\qw	&	\gate{Q_4}	&	\qw	\qw	\\
    }

}\end{center}
In the circuit, $C_1$ and $C_2$ combine the terms with the ratio of the coefficients $\alpha$ and $\beta$.    
Note that if $A$ is approximated by using $r$ terms, then the whole circuit would require $O(rn)$ quantum gates.

\section{Conclusion}
In this paper, we study the approximation of a given matrix or a vector by using its tensor decomposition. 
We show the results for the example distributions, machine learning datasets, variational quantum circuits, and Ising-type quantum Hamiltonians.
The method can be used to map data matrices into quantum states and also can be used to approximate operators to design more efficient quantum circuits, which is particularly useful for the simulation of quantum systems on quantum computers.
The method can be also used to approximate quantum operations when doing their classical simulations.
Note that this approximation may not be a good way to make a single computation since it requires many Schmidt decomposition. However, it can be a good way to reduce the computations for frequently used tools such as QFT or to simulate some dynamics of quantum systems, where the number of terms is fixed but the parameters that construct the term change over time. 
In addition, it can be used in the training of machine learning tools to reduce the size of a trained model.

\section{Acknowledgement}
S.K. would like  to acknowledge the financial support from the National Science Foundation under Award No. 1955907 and the support of the U.S. Department of Energy (Office of Basic Energy Sciences) under Award No. DE-SC0019215. 

\bibliography{main}

\end{document}